\documentclass[prl,aps,floats,letterpaper,floatfix,superscriptaddress,preprintnumbers,twocolumn]{revtex4}
\usepackage{amsmath,color,ulem}
\usepackage{amssymb}
\usepackage[dvips]{graphicx}
\definecolor{darkblue}{rgb}{0,0,0.5}
\usepackage[dvips, unicode, colorlinks, bookmarks=true, citecolor=darkblue, linkcolor=darkblue, urlcolor=darkblue]{hyperref}

\newcommand{\comment}[1]{}

\newcommand{\rem}[1]{}
\begin{document}
\title{Creation of a quantum oscillator by classical control }
\author{Stefan Danilishin}
\affiliation{Max-Planck Institut f\"ur Gravitationsphysik (Albert-Einstein-Institut) and Leibniz Universit\"at Hannover, Callinstr. 38, 30167 Hannover, Germany}
\affiliation{Physics Faculty, Moscow State University, Moscow 119991, Russia}
\author{Helge M\"uller-Ebhardt}
\author{Henning Rehbein}
\affiliation{Max-Planck Institut f\"ur Gravitationsphysik (Albert-Einstein-Institut) and Leibniz Universit\"at Hannover, Callinstr. 38, 30167 Hannover, Germany}
\author{Kentaro Somiya}
\affiliation{Theoretical Astrophysics 130-33, California Institute of Technology, Pasadena, CA 91125, USA}
\affiliation{Max-Planck Institut f\"ur Gravitationsphysik (Albert-Einstein-Institut), Am M\"uhlenberg 1, 14476 Golm, Germany}
\author{Roman Schnabel}
\author{Karsten Danzmann}
\affiliation{Max-Planck Institut f\"ur Gravitationsphysik (Albert-Einstein-Institut) and Leibniz Universit\"at Hannover, Callinstr. 38, 30167 Hannover, Germany}
% \affiliation{Max-Planck Institute for Gravitational Physics and University of Hannover, Callinstr 38, 30167, Hannover, Germany}
\author{Thomas Corbitt}
\author{Christopher Wipf}
\author{Nergis Mavalvala}
\affiliation{LIGO Laboratory, NW22-295, Massachusetts Institute of
Technology, Cambridge, MA 02139, USA}
\author{Yanbei Chen}
\affiliation{Theoretical Astrophysics 130-33, California Institute of Technology, Pasadena, CA 91125, USA}
\affiliation{Max-Planck Institut f\"ur Gravitationsphysik (Albert-Einstein-Institut), Am M\"uhlenberg 1, 14476 Golm, Germany}

\begin{abstract}
  As a pure quantum state is being approached via linear feedback, and
  the occupation number approaches and eventually goes below unity,
  optimal control becomes crucial.  We obtain theoretically the
  optimal feedback controller that minimizes the uncertainty for a
  general linear measurement process, and show that even in the
  absence of classical noise, a pure quantum state is not always
  achievable via feedback.  For Markovian measurements, the deviation
  from minimum Heisenberg Uncertainty is found to be closely related
  to the extent to which the device beats the free-mass Standard
  Quantum Limit for force measurement.  We then specialize to optical
  Markovian measurements, and demonstrate that a slight modification
  to the usual input-output scheme --- either injecting frequency
  independent squeezed vacuum or making a homodyne detection at a
  non-phase quadrature --- allows controlled states of kilogram-scale
  mirrors in future LIGO interferometers to reach occupation numbers
  significantly below unity.
\end{abstract}

\maketitle

{\bf Motivation.} The detectors of the Laser Interferometer
Gravitational-wave Observatory (LIGO)~\cite{LIGO} are currently
operating at a factor of 10 above the Standard Quantum Limit
(SQL)~\cite{BK92} at closest approach, limited by classical noise
below about 100 Hz, and by quantum shot noise above 100
Hz~\cite{S5LIGO}. This low noise performance allows for probing and
manipulating LIGO mirrors (10\,kg each and suspended as pendulums with
resonant frequency $\omega_p = 2 \pi\times 0.7$~Hz and typical damping
rate $\gamma_p \sim 10^{-6} {\rm \,sec}^{-1}$) close to scales set by
the Heisenberg Uncertainty Principle.  Recently, electronic feedback
control was used to shift the resonant frequency of the pendulum mode
up to $\sim140$ Hz and damp it, leading to an effective occupation
number of 234 for the kg-scale
oscillator~\cite{LSCcool}. Semiclassical calculations estimate an
effective occupation number of $N_{\rm eff} \approx Q_{\rm eff}
{S_{x}(\Omega_{\rm eff})}/{S_{\rm SQL}(\Omega_{\rm eff})}$, where
$S_x$ is the detector's position-referred noise spectral density,
$S_{\rm SQL} = 2\hbar/(m\Omega^2)$ is the free-mass SQL for position
measurement, $\Omega_{\rm eff}$ is the eigenfrequency of the
controlled oscillator, and $Q_{\rm eff}$ is its quality factor.
Extrapolation to $N_{\rm eff} <1$ favors $Q_{\rm eff} \ll 1$; yet this
is exactly where the approximation fails, and a more careful treatment
of optimal control is required.  Moreover, we must answer the
following question: {\it does the strong continuous position
  measurement required for making the stiff electro-optical potential
  always produce significant decoherence of the oscillator's quantum
  state?}

The cold damping technique used in Ref.~\cite{LSCcool} was first
proposed by Mancini et al.~\cite{manciniPRL1998} and demonstrated
experimentally by Cohadon et al.~\cite{cohadonPRL1999}. Subsequent
experiments have used cold damping to cool mechanical oscillators with
the goal of reaching the quantum ground state
\cite{cohadonPRL1999,courtyEuroPhys2001,klecknerNature2006,corbittPRL2007,
  poggioPRL2007,vivantePRL2008,mowlowryPRL2008,LSCcool}, typically
using simple (proportional plus derivative) filters.
% (constant plus derivative -- or PD control). 
Little attention has been paid to the question of whether these
filters are optimal.  Furthermore, feedback control was used only for
damping in these experiments, and not for shifting the resonant
frequency of the oscillator, as in the LIGO experiment~\cite{LSCcool}.
In this Letter, we obtain the optimal state-preparation control scheme
for a general linear measurement process (possibly non-Markovian), and
study prospects of quantum-state preparation for kg-scale test masses
in future LIGO detectors via feedback control.  Our general theory
also applies to other mechanical
structures~\cite{cohadonPRL1999,courtyEuroPhys2001,klecknerNature2006,corbittPRL2007,
  poggioPRL2007,vivantePRL2008,mowlowryPRL2008}.

It turns out that occupation number is not necessarily a good
measure of ``quantum-ness'': squeezed states, for example, can have
high occupation number, yet they are more ``quantum'' than a vacuum
state. Moreover, the definition of occupation number requires a
well-defined real-valued eigenfrequency, which can be ambiguous for
two reasons: (i) the controller can modify the oscillator's original
eigenfrequency $\omega_p$; and (ii) for oscillators with a finite
quality factor $Q_{\rm eff}$, the choice for an effective real
eigenfrequency $\Omega_{\rm eff}$ would be ambiguous by $\sim
\Omega_{\rm eff}/Q_{\rm eff}$. Instead we use the {\it purity},
defined as
\begin{equation}
U \equiv \sqrt{V_{xx} V_{pp} - V_{xp}^2} \ge \hbar/2.
\end{equation}
where $V_{xx}$, $V_{pp}$ and $V_{xp} $ are uncertainties of
oscillator's position and momentum along with their
cross-correlation. For steady states, which have $V_{ xp}=0$, $U$ can
be converted to an effective occupation number,
\begin{equation}\label{Neffdef}
N_{\rm eff} \equiv  U/\hbar-1/2\,,
\end{equation}
which is the {\it minimum} occupation number one could obtain when the
same mirror state is put into a quadratic potential well with an
arbitrary eigenfrequency $\Omega$:
\begin{equation}
\label{Neff}
N_{\rm eff} +1/2 =\min_{\Omega}\left[\left({V_{ pp}/(2m) + m\Omega^2 V_{ xx}/2}\right)/({\hbar\Omega})\right],
\end{equation}
achieved at $\Omega = \sqrt{V_{pp}/(m^2V_{xx})}$. Since $\Omega$ may
be very different from the oscillator eigenfrequency, the resulting
quantum state tends to be {\it position squeezed} if $\Omega >
\omega_p$, and {\it momentum squeezed} if $\Omega
<\omega_p$~\footnote{In
  Refs.~\cite{cohadonPRL1999,courtyEuroPhys2001,klecknerNature2006,corbittPRL2007,
    poggioPRL2007,vivantePRL2008,mowlowryPRL2008,LSCcool}, either
  kinetic or total energy is compared with quanta of the original
  oscillator; this will become ambiguous when $Q_{\rm eff}$ becomes
  low, and irrelevant when $\Omega_{\rm eff}$ is shifted
  significantly.}.  $N_{\rm eff}$ also determines the von Neumann
entropy of the state: $ S = (N_{\rm eff}+1)\log(N_{\rm eff}+1) -
N_{\rm eff} \log N_{\rm eff}$.

\begin{figure}[t]
  \includegraphics[width=.35\textwidth]{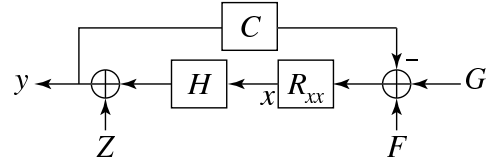}
  \vspace{-0.25cm} \caption{Block diagram of the feedback control
    system. } \label{fig1}
\end{figure}

{\bf General Optimal Controller.}  A block diagram of the entire
measurement-control system is shown in Fig.~\ref{fig1}, where $x$ is
the position of the oscillator, $y$ is the output field we measure,
$H$ is the measurement transfer function, $\tilde R_{xx} =
-1/[(\Omega-\omega_p + i\gamma_p)(\Omega+\omega_p + i\gamma_p)] $ is
the response function of the oscillator (with $\gamma_p =
\omega_p/Q_p$ its relaxation rate; the tilde denotes a
frequency-domain quantity), $F$ and $Z$ are force and sensing noises,
$G$ is a possible classical force acting on the oscillator, and $C$ is
the feedback kernel.  We write the closed-loop position and momentum
of the oscillator as
\begin{equation}
\label{ct}
\tilde x_{\rm ctrl}= \tilde x_0  - \tilde K_{\rm ctrl}\tilde y_0 \,,\quad \tilde p_{\rm ctrl} = \tilde p_0 + i  \Omega  \tilde  K_{\rm ctrl}\tilde y_0
\end{equation}
where $x_0$, $p_0$ are the open-loop evolution of the oscillator
position and momentum (we set its mass equal to $1$, and use $\tilde p
= -i\Omega\tilde x$), $y_0$ is the open-loop out-going field, and
\begin{equation}
\label{kc} \tilde K_{\rm ctrl} =  \tilde R_{xx} \tilde C/({1 +
\tilde  R_{xx} \tilde C \tilde H})\,.
\end{equation}

The closed-loop dynamics is {\it stable} and the feedback $\tilde C$
{\it proper,} {\it if and only if} (i) $\tilde K_{\rm ctrl}$ is causal
(i.e., no poles in the upper-half complex plane); and (ii)
$\lim_{\Omega\rightarrow \infty} \Omega \tilde K_{\rm ctrl}(\Omega)
=0$.  The closed-loop response of the oscillator's position to
external force is given by
\begin{equation}\label{succepeqn}
  \tilde R_{xx}^{\rm eff} =\tilde R_{xx}(1-\tilde H \tilde K_{\rm ctrl})\,.
\end{equation}
In Eq.~\eqref{ct}, closed-loop quantities are viewed as subtracting
the open-loop readout field from open-loop quantities.  On the other
hand, causal Wiener filters $\tilde K_{a}$ ($a = x,p$) can be
constructed based on the cross spectral density between $x_0$, $p_0$
and $y_0$, to yield the best (least-mean-square) estimates of $x_0$
and $p_0$ based on past measurement of $y_0$ \cite{PapoulisBook}. In
terms of these filters, we have
\begin{eqnarray}
\left[
\begin{array}{c}
V_{xx}^{\rm ctrl} \\
V_{pp}^{\rm ctrl}
\end{array}
\right] 
= 
\left[
\begin{array}{c}
V_{xx}^c \\
V_{pp}^c
\end{array}
\right] 
+ \int_0^{\infty}  \frac{d\Omega}{2\pi} 
\left[
\begin{array}{c}
|\tilde K_{\rm ctrl}- \tilde K_{x}|^2 \\
|i\Omega \tilde K_{\rm ctrl}- \tilde K_{p}|^2
\end{array}
\right]  S_{yy} \,,
\end{eqnarray}
where $S_{yy}$ is the single-sided spectral density of
$y_0$~\footnote{$V_{ab}^c$ and $V_{ab}^{\rm ctrl}$ ($a,b=x,p$) stand
  for uncertainties of conditional and controlled states
  accordingly.}.  Minimizing over all stable closed-loop systems, we
obtain
\begin{equation}
  \label{opt} U_{\rm opt}\equiv \min_{K_{\rm ctrl}} \sqrt{V_{xx}^{\rm
      ctrl}V_{pp}^{\rm ctrl}} =  \sqrt{V_{xx}^{\rm c} V_{pp}^{\rm c}}+
  V_{xp}^{\rm c}\,,
\end{equation}
which is achieved by a unique controller with
\begin{equation}
\label{Kctrl}
\tilde K_{\rm ctrl} = \frac{1}{ \tilde \phi_+(\Omega)} \left[ \tilde
  G_x(\Omega) - \frac{G_x(0)}{\rho - i\Omega}\right],\,\rho=\sqrt{\frac{V_{pp}^{\rm
      c}}{V_{xx}^{\rm c}}}\,.
\end{equation}
Here $\tilde\phi_+^{-1}(\Omega)$ is a {\it causal whitening filter}
for the output field $y_0$, with ${\tilde\phi}_+^* \tilde \phi_+
=S_{yy} $ and both $\tilde \phi_+$ and $\tilde \phi_+^{-1}$ analytic
in the upper-half complex plane, while $\tilde G_x \equiv
[S_{xy}/\phi_+^*]_+$, where $S_{xy}$ is the cross spectral density
between $x_0$ and $y_0$, and $[\tilde F]_+$ denotes {\it extracting
  the causal part of} $\tilde F$ while $F^*$ stands for the complex
conjugate of $F$. More specifically, under the scaling transform
$\{x',p'\}=\{x/\sqrt{V_{xx}^c},p/\sqrt{V_{pp}^c}\}$, the error ellipse
of the optimally controlled state becomes a circle with $V_{x'x'}^{\rm
  ctrl} = V_{p'p'}^{\rm ctrl} = 1+V_{x'p'}^c$, which is in turn equal
to the larger eigenvalue of the re-scaled conditional covariance
matrix. Thus the error ellipse of the optimally controlled state is
the one with the minimum area among those that (i) totally encompass
the conditional-state error ellipse; and (ii) are consistent with
$V_{xp}^{\rm ctrl}=0$. This means the controlled state is always a
mixed state, unless $V_{xp}^c=0$~\footnote{$V_{xp}=0$ corresponds to
  $G_x(0)=0$, which in turn requires that no information about $x(t)$
  be collected at time $t-0$. This point will be elaborated in a
  future publication.}.

{\bf General Markovian measurements.} We consider the open-loop system
(Fig.~\ref{fig1})
\begin{equation}
\label{mark}
y = Z +x\,,\quad  x = \tilde R_{xx}(F+G)
\end{equation}
where we have set $H=1$ (without loss of generality), and $Z$ and $F$
are now characterized by {\it real and constant} (single-sided) cross
spectral densities $S_{ZZ}$, $S_{ZF}$ and $S_{FF}$ satisfying the
Heisenberg Uncertainty Relation~\cite{BK92}
\begin{equation}
  \label{heisenberg1} \sqrt{S_{ZZ} S_{FF} -S_{ZF}^2} \equiv  \mu \hbar\,,
\end{equation}
where $\mu \ge 1$ measures the {\it purity} of the measurement process
(with $\mu=1$ corresponding to quantum-limited measurement).  This
describes measurement systems with white sensing and force noises,
e.g., measurement of mirror location using a broadband Fabry-Perot
cavity, with frequency independent input squeezing and homodyne
detection. We recast $S_{yy}$ into a {\it causally factorized form}:
\begin{eqnarray}
\label{Syy}
S_{yy} &=&S_{ZZ} +2 \mathrm{Re}(\tilde R_{xx}) S_{ZF} + S_{FF} |\tilde R_{xx}|^2 \nonumber \\
&\equiv & S_{ZZ} L L^*/(P P^*)\,.
\end{eqnarray}
Here $P \equiv -1/\tilde R_{xx}$, $L L^*\equiv \Omega^4- 2A \omega_p^2
\Omega^2 + B^2 \omega_p^4 $,
\begin{equation}
\label{AB}
A \equiv 1+ \frac{1}{\omega_p^2}\frac{S_{ZF}}{ S_{ZZ}}\,,
\quad
B^2 \equiv 1 + \frac{2}{\omega_p^2}\frac{S_{ZF}}{S_{ZZ}}
+\frac{1}{\omega_p^4} \frac{S_{FF}}{ S_{ZZ}}\,.
\end{equation}
$L$ is defined here in such a way that it only has roots in the upper-half complex plane~\footnote{We
  have set $\gamma_p \rightarrow 0$ in the definition of $A$ and
  $B$.}. Using the definition for $\tilde K_a$ and Eqs.~\eqref{Syy},
\eqref{AB}, the conditional covariance matrix can be obtained:
\begin{equation}
\mathbf{V} = \frac{\hbar\mu}{2}
\left[
\begin{array}{cc}
\frac{1}{\omega_p}\sqrt{\frac{2}{A+B}} &  \sqrt{\frac{B-A}{B+A}} \\
\sqrt{\frac{B-A}{B+A}} &  \omega_p\sqrt{\frac{2B^2}{A+B}}
\end{array}\right]
\end{equation}
The conditional purity is given by
\begin{equation}
\label{cond_purity}
U_c  = \sqrt{|\mathbf{V}|} = \mu \hbar/2\,,
\end{equation}
which is {\it identical} to the ``purity'' of the measurement
process; see Eq.~\eqref{heisenberg1}. In the absence of classical
noise, the conditional quantum state of the oscillator is always
pure. With Eq.~\eqref{opt}, we obtain
\begin{equation}
  \label{Uctrl}
  {U_{\rm ctrl}}/({\hbar/2}) =\mu ({\sqrt{1-A/B}+\sqrt{2}})/{\sqrt{1+A/B}}
\end{equation}
which is achieved by the unique optimal filter [cf.~\eqref{kc} and
\eqref{Kctrl}] with associated closed-loop response function
\begin{eqnarray}\label{ctrl}
  \tilde C \!\! &=& \!\! C_0(\Omega - C_1)/(\Omega-C_2)\,, \\
  \tilde R_{xx}^{\rm eff}\!\! &=& \!\! -
  (\Omega-\Omega_4)/[(\Omega-\Omega_1)(\Omega-\Omega_2)(\Omega-\Omega_3)],\quad
\end{eqnarray}
where $C_0 = -(\omega_p^2 + \Omega_4\Omega_3)$, $C_1 =
(\Omega_3^3+\omega_p^2\Omega_4)/(\omega_p^2+\Omega_4\Omega_3)$, $C_2 =
\Omega_4$ and $\Omega_{1,2}$ are roots of $Q$, namely
\begin{equation}
\label{omega12}
\Omega_{1,2} = \pm \omega_p \sqrt{({B+A})/2}-i  \omega_p  \sqrt{({B-A})/{2}}
\end{equation}
while $\Omega_{3,4}$ are purely imaginary:
\begin{equation}
  \Omega_3 = -i \sqrt{B}\omega_p\,,\quad
  \Omega_4 = -i [\sqrt{B}+\sqrt{2(B-A)}]\omega_p\,.
\end{equation}
We note that: (i) the poles of the closed-loop dynamics,
$\Omega_{1,2}$, are identical to the zeros of $S_{yy}$, i.e., the
optimal controller ``finds'' the frequency of maximal sensitivity, and
shifts the oscillator's eigenfrequency there; (ii) the $\Omega$ that
achieves $N_{\rm eff}$ in Eq.~\eqref{Neff} and motivates $N_{\rm eff}$
as an occupation number in a harmonic potential, is equal to
$|\Omega_{1,2}|$, the modulus of the closed-loop poles; (iii) the
optimal controller \eqref{ctrl} is in fact proportional feedback plus
constant damping and simple band limiting (which is required for
$V_{pp}^{\rm ctrl}$ to be finite), justifying previous choices
\cite{cohadonPRL1999,courtyEuroPhys2001,klecknerNature2006,
  poggioPRL2007,corbittPRL2007,vivantePRL2008,mowlowryPRL2008,LSCcool};
and (iv) a pure state is only strictly achievable when $\mu\approx 1$
(i.e., in the absence of classical noise) {\it and} $A\approx B$ (in
general $|A|\le B$).

Having $A\approx B$ corresponds to a high quality factor for the
closed-loop dynamics [Cf.~Eq.~\eqref{omega12}], with
\begin{equation}
  Q_{\rm eff} = {\sqrt{B+A}}/{(2\sqrt{B-A})}.
\end{equation}
This is consistent with our understanding that a low-$Q$ oscillator
cannot have a pure quantum state due to the Fluctuation Dissipation
Theorem~\cite{BK92}. Moreover, $A\approx B$ also corresponds to
$S_{yy}$ having a very small minimum [in the limit of $A=B$, $S_{yy}$
reaches 0; cf.~Eq.~\eqref{Syy}]. Let us consider the force
($G$)-referred noise spectrum,
\begin{equation}
  S_G = S_{yy}/|\tilde R_{xx}|^2 = S_{ZZ} L L^*,
\end{equation}
and compare it with the free-mass SQL for force detection $S_G^{\rm
  SQL} = 2 \Omega^2 \hbar$. Taking the minimum over all frequencies,
we obtain the factor by which the SQL is beaten:
\begin{equation}
  \eta^2\equiv [S_G/S_G^{\rm SQL}]_{\rm min} = \mu/(2{Q_{\rm eff}})
\end{equation}
which leads to [cf.~Eq.~\eqref{Uctrl}]
\begin{equation}
\label{bound}
\frac{{U}_{\rm ctrl}}{\hbar/2} = \eta^2 + \frac{\sqrt{2} \mu}{\sqrt{1 + A/B}} \ge  \eta^2 + 1,\; N_{\rm eff} \ge\eta^2/2.
\end{equation}
This means an oscillator under measurement can only be converted into
a quantum oscillator via control if it can beat the free-mass SQL in a
force measurement, in which case the optimally-controlled closed-loop
quality factor $Q_{\rm eff}$ would also far exceed unity.

{\bf Position measurement with light.} In the realistic case with
suspension and internal thermal noises and optical loss, we have (as
in Ref.~\cite{HelgePRL2008})
\begin{subequations}\label{eom}
\begin{eqnarray}
  x &=& \tilde R_{xx} [ \alpha a_1 + \xi_F +G] \\
  y &=& a_1 \sin\phi + \cos\phi [a_2+\alpha/\hbar(x+\xi_x)] +\sqrt{\epsilon} n \quad
\end{eqnarray}
\end{subequations}
or in the notation of Eq.~\eqref{mark}:
\begin{eqnarray}
 F = {\alpha a_1 + \xi_F},\;
 Z=  \xi_x +\frac{a_1 \sin\phi + a_2 \cos\phi  +\sqrt{\epsilon} n}{\alpha/\hbar \cos\phi},\;
\end{eqnarray}
where $a_{1,2}$ are the input quadrature fields, $\phi$ is the readout
phase (0 for phase quadrature and $\pi/2$ for amplitude quadrature),
{$\alpha$} is the measurement strength ($\alpha= 4\sqrt{\hbar\omega_0
  I_c/(\tau c^2)}$ for a Michelson interferometer with arm cavities,
with $\omega_0$ the carrier frequency, $I_c$ the circulating power in
the arms, $\tau$ the input-mirror power transmissivity, $c$ the speed
of light), $\epsilon$ is the optical loss, and $n$ is the vacuum
noise. $\xi_x$ and $\xi_F$ are the classical sensing and force noises,
respectively, whose spectra cross the position- and force-SQL at
frequencies $\Omega_q /\zeta_x$ and $\Omega_q \zeta_F$, respectively,
where $\Omega_q \equiv \alpha/\sqrt{\hbar}$ is the characteristic
measurement frequency (as defined in Ref.~\cite{HelgePRL2008}).

\begin{figure}[t]
  \includegraphics[height=.22\textwidth]{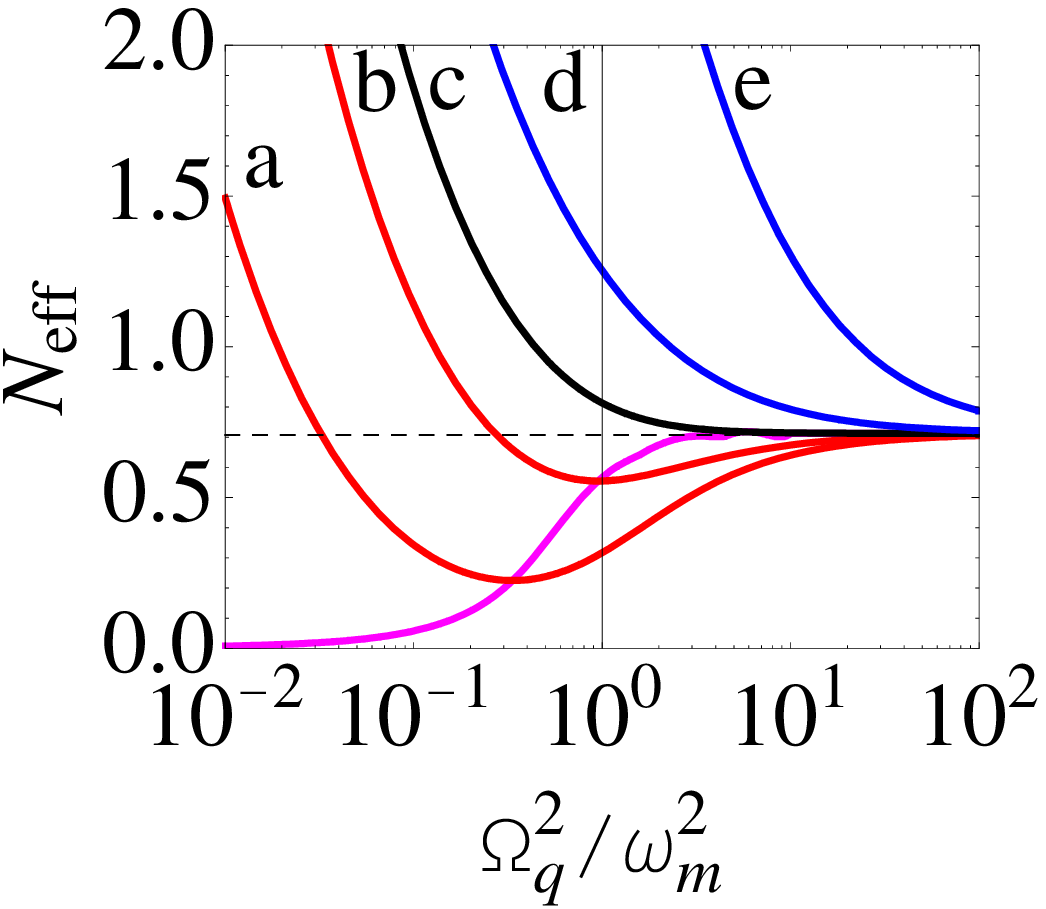}\includegraphics[height=0.22\textwidth]{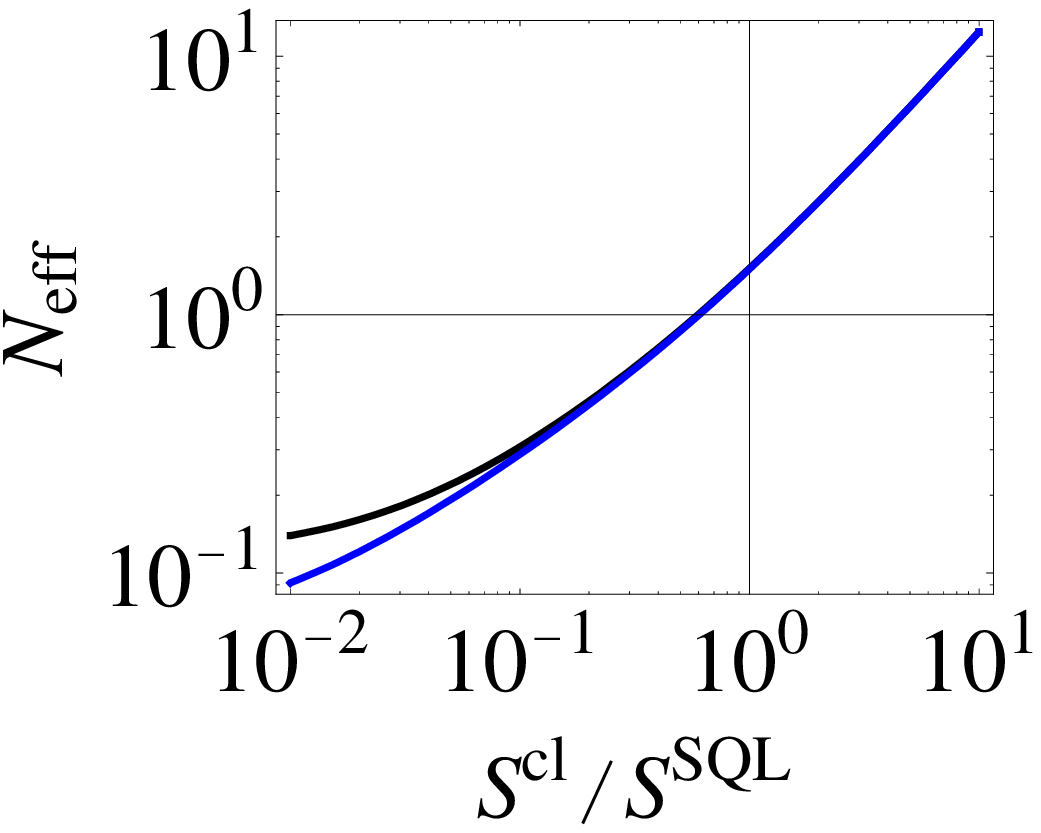}\\
  \caption{\textit{Left panel:} $N_{\rm eff}$ as a function of
    measurement strength $\Omega_q^2/\omega_p^2$. Curves (a-e)
    represent $N_{\rm eff}$ at different temperatures $\theta$: a)
    $0.1$, b) $0.5$, c) $1$ (critical), d) $2$, e) $10$. \textit{Right
      panel:} Minimal $N_{\rm eff}$ as a function of factor $\eta_{\rm
      cl}^2$ for $10$~dB squeezing (blue) and vacuum input
    (black).\label{fig2}}
\end{figure}

{\bf Viscosity noise alone and phase-quadrature readout.} Here we
consider the cold damped systems of Refs.
\cite{cohadonPRL1999,courtyEuroPhys2001,klecknerNature2006,poggioPRL2007,corbittPRL2007,vivantePRL2008,mowlowryPRL2008,LSCcool},
i.e. $\epsilon=\zeta_x=\phi=0$, $\omega_p \neq 0$, $\zeta_F^2 = 4
\gamma_p k_B T/(\Omega_q^2 \hbar)$. Since a free mass under such
measurements never beats the free-mass SQL, according to
Eq.~\eqref{bound}, it will always have $N_{\rm eff} \ge 1/2$.  In
order to have a lower $N_{\rm eff}$, we have to rely on beating the
free-mass SQL around the oscillator's original resonant frequency,
which is only possible when $k_B T/(\hbar \omega_p Q_p) <1/2$.
Analytic results reveal a phase-transition-like situation: below a
{\it critical temperature}, with
\begin{equation}
  \theta \equiv T/T_c <1, \quad T_c \equiv \hbar\omega_p Q_p/(2\sqrt{2}k_B)\,,
\end{equation}
the minimum occupation number
\begin{equation}N_{\rm opt}(\theta) =
  2^{-3/2}[\sqrt{2-\theta^2}+\sqrt{2\theta\sqrt{2-\theta^2}}+\theta-\sqrt{2}]
  % \frac{\sqrt{2-\theta^2}+\sqrt{2\theta\sqrt{2-\theta^2}}+\theta-\sqrt{2}}{2\sqrt{2}}\,,
\end{equation}
$N_{\rm opt} \sim 2^{-3/4}\theta^{1/2}$ as $\theta\rightarrow 0$, can
be achieved with
\begin{equation}
  \frac{\Omega_q}{\omega_p} =
  \sqrt{\frac{\theta^{1/2}(2-\theta^2)^{3/4}}{\sqrt{2-\theta^2}-\theta}
    -\frac{\theta}{\sqrt{2}}} \sim
  (\sqrt{2}\theta)^{1/4}\,,\;(\theta\rightarrow 0),
\end{equation}
while for $T>T_c$, a temperature-independent minimum of $N_{\rm eff} =
1/\sqrt{2}$ is achieved with infinite measurement strength as
indicated in Fig.~\ref{fig2} (left panel).

For a LIGO detector, $T \gg T_c$, so $N_{\rm eff} \ge 1/\sqrt{2}$ even
if viscous damping alone is considered. For the systems of
Refs.~\cite{cohadonPRL1999,courtyEuroPhys2001,klecknerNature2006,
  poggioPRL2007}, prospects for surpassing
%   \begin{equation}
$ T_c \simeq 17\,{\rm K}\times (Q_p/10^6) \times [\omega_p/(2\pi\times
1\,{\rm MHz})]$
%   \end{equation}
are much more promising.

{\bf Prospects for LIGO ($\omega_p=0$). } In order to evade the
limitation of $N_{\rm eff} \ge 1/\sqrt{2}$, we consider an arbitrary
readout quadrature with squeezed light input. Given a classical noise
budget, with vacuum input or fixed squeezing factor (10\,dB), we
optimize $\Omega_q$, $\phi$ and the squeezing angle for a minimum
$N_{\rm eff}$. In the right panel of Fig.~\ref{fig2} , we plot $N_{\rm
  eff}$ as a function of the factor $\eta_{\rm cl}^2 \equiv [S^{\rm
  cl}/S^{\rm SQL}]_{\rm min} = 2\zeta_F\zeta_x$ by which the total
classical noise beats the SQL~\cite{HelgePRL2008}, while fixing the
optical loss $\epsilon=0.01$. The input squeezing appears to be not so
crucial and almost the same results can be achieved without it. A
detailed optimization will be presented elsewhere.

{\bf Summary} In this Letter, we developed the general theory for
optimal state-preparation via linear feedback control. The optimally
controlled $x$-$p$ error ellipse, achievable using the {\it unique}
optimal controller, is the one with minimum area that still maintains
$V_{xp}=0$ and encompasses the conditional-state error ellipse.  For a
general Markovian measurement process, the conditional-state purity
equals that of the measurement [cf.~Eq.~\eqref{cond_purity}], and the
absence of classical noise guarantees a pure conditional state; yet a
nearly pure controlled state requires additionally that correlations
exist between sensing and force noises, in such a way that the device
beats the free-mass SQL significantly in a force measurement. In this
case, the optimal controller creates an oscillator with an
eigenfrequency $\Omega_{\rm eff}$ around the most sensitive frequency
of the device, with quality factor $Q_{\rm eff} \gg 1$ and an
effective occupation number $N_{\rm eff} \ll 1$. This $N_{\rm eff}$ is
also meaningfully measured against a harmonic potential with
real-valued eigenfrequency $\Omega \approx \Omega_{\rm
  eff}$. Furthermore, restricting to conventional measurements applied
to a viscously damped oscillator, we found a critical temperature
$T_c$, above which the oscillator is limited by $N_{\rm eff} \ge
1/\sqrt2$, while below which, $N_{\rm eff}$ approaches 0 as $T/T_c
\rightarrow 0$. Finally, for LIGO, the answer to the question raised
at the beginning of this Letter is ``yes'' for conventional
phase-quadrature readout, where the need to up-shift the oscillator's
eigenfrequency does limit $N_{\rm eff}$ above $1/\sqrt{2}$, and ``no''
for an optimal choice of measurement strength and (non-phase) readout
quadrature (possibly in combination with input squeezing), where the
sub-SQL classical noise budget of future detectors will allow an
optimally controlled kg-scale oscillator to achieve $N_{\rm eff}\ll
1$.

{\bf Acknowledgment} We thank the AEI-Caltech-MIT-MSU MQM group for
many discussions. Research of S.D., H.M-E., K.S. and Y.C. is supported
by the Alexander von Humboldt Foundation.  S.D., Y.C.\ and K.S.\ are
also supported by the National Science Foundation (NSF) grants
PHY-0653653 and PHY-0601459, as well as the David and Barbara Groce
startup fund at Caltech. K.S. is supported by Japan Society for the
Promotion of Science. T.C, N.M. and C.W. are supported by NSF grants
PHY-0107417 and PHY-0457264, and by the Sloan Foundation.

\end{document}